\documentstyle[12pt]{article}
\newcommand{\II}{{\cal I}}
\newcommand{\MM}{{\cal M}}

\newcommand{\wt}{\widetilde}

\newcommand{\be}{\begin{equation}}
\newcommand{\ee}{\end{equation}}
\newcommand{\ben}{\begin{eqnarray}\displaystyle}
\newcommand{\een}{\end{eqnarray}}
\newcommand{\refb}[1]{(\ref{#1})}
\newcommand{\p}{\partial}
\newcommand{\sectiono}[1]{\section{#1}\setcounter{equation}{0}}

\begin{document}

{}~ \hfill\vbox{\hbox{hep-th/9702061}\hbox{MRI-PHY/97-5}
\hbox{NI 97006}}\break

\vskip 3.5cm

\centerline{\large \bf F-theory and the Gimon-Polchinski Orientifold}

\vspace*{6.0ex}

\centerline{\large \rm Ashoke Sen\footnote{On leave of absence from 
Tata Institute of Fundamental Research, Homi Bhabha Road, 
Bombay 400005, INDIA}
\footnote{E-mail: sen@mri.ernet.in, sen@theory.tifr.res.in}}

\vspace*{1.5ex}

\centerline{\large \it Mehta Research Institute of Mathematics}
 \centerline{\large \it and Mathematical Physics}

\centerline{\large \it  Chhatnag Road, Jhoosi,
Allahabad 221506, INDIA}
\vspace*{1.5ex}

\centerline{\large \it and}
\vspace*{1.5ex}

\centerline{\large \it Issac Newton Institute for Mathematical Sciences}
\centerline{\large \it University of Cambridge}

\centerline{\large \it Cambridge, CB3 0EH, U.K.} 

\vspace*{4.5ex}

\centerline {\bf Abstract}

We establish the equivalence of the Gimon-Polchinski orientifold 
and F-theory on
an elliptically fibered Calabi-Yau three fold on base 
$CP^1 \times CP^1$ by
comparing the gauge symmetry breaking pattern, local deformations
in the moduli space, as well as the axion-dilaton background in the weak
coupling limit in the two theories.
We also provide an explanation for an apparent discrepancy between the
F-theory and the orientifold results for constant coupling configuration.

\vfill \eject

\baselineskip=18pt

\sectiono{Introduction } \label{s1}

Conventional compactification of type IIB string theory is characterized by
the property that both, the dilaton and the scalar field arising in the 
Ramond-Ramond (RR) sector of the theory (which we shall refer to as the
axion), are constant on the internal space.
F-theory\cite{VAFAF,FTHEORY} 
provides us with a novel way of compactifying type IIB string theory
that does not suffer from this restriction. 
The starting point in an F-theory compactification is a manifold $\MM$
that admits elliptic fibration over a base $B$. F-theory on $\MM$ is by
definition type IIB on $B$, with the axion-dilaton modulus set equal to the
complex structure modulus of the fiber torus at every point on the
base.
In a generic F-theory compactification the scalar fields undergo
non-trivial SL(2,Z) monodromy around closed cycles of the internal manifold. 
Since
SL(2,Z) is a non-perturbative symmetry of the type IIB theory, there is no
conventional perturbative description of F-theory compactification.

It was conjectured  by Vafa\cite{VAFAF}
that F-theory on a K3 surface elliptically fibered
over a base $CP^1$ is dual to heterotic string theory on $T^2$.
This was established in \cite{SENF} by examining the F-theory background at
a special point in the moduli space corresponding to the orbifold limit of 
K3, where the dilaton-axion field becomes constant on the base, and
hence the theory describes a conventional string compactification. This
was found to be an orientifold\cite{ORIENT,DBRANE}
of type IIB theory on $T^2$, which is
related by T-duality to type I on $T^2$. The conjectured 
equivalence between type
I and the $SO(32)$ heterotic string theory in ten dimensions\cite{WITTEND}
then establishes the
duality between F-theory on K3 and heterotic string theory on $T^2$.

It was also shown in \cite{SENF} that away from this special point
in the moduli space, the axion-dilaton background in the F-theory agrees
with that in the orientifold theory in the weak coupling limit, but 
they differ by non-perturbative terms. This was interpreted as due to
the quantum corrections to the orientifold background which modifies it
to the F-theory background. This was proved in \cite{BDS} by using a
three brane to probe the orientifold background.

In this paper we shall carry out a similar analysis for F-theory on a
Calabi-Yau 3-fold 
with Hodge numbers $(h_{11}=3, h_{12}=243)$ which admits an elliptic 
fibration over the base $CP^1\times CP^1$\cite{VAFAF,FTHEORY}.
(Some aspect of this compactification has recently been discussed in
ref.\cite{JATKAR} following ref.\cite{DASMUK}.)
Thus this describes a compactification of type IIB theory on $CP^1
\times CP^1$ with varying axion-dilaton field. 
This model has been conjectured\cite{FTHEORY} to be dual to $E_8\times E_8$
heterotic string theory on K3, with the total instanton number 24 equally
divided among the two $E_8$ subgroups. This, in turn, has been 
conjectured\cite{WITSIX} to
be dual to an orientifold of type IIB 
theory compactified on $T^4$, constructed
by Gimon and Polchinski\cite{GIMPOL} (see also refs.\cite{BIAN} for earlier
construction of this model at special points in the moduli space).
The main purpose of this paper will be to establish the duality between this
particular F-theory compactification and a T-dual of the 
Gimon-Polchinski (GP) model
directly following methods similar to that in \cite{SENF}. Some attempts
in this direction were made earlier in refs.\cite{GIMJOH,GOPMUK}, and
duality between a different pair of
orientifold and F-theory vacua in six dimensions
was established in refs.\cite{BLUZAF,DABPAR}.

The particular T-dual of the GP model that we shall consider may
be described as type IIB on $CP^1\times CP^1$ with a configuration of
four orientifold seven planes and eight pairs of Dirichlet seven branes
transverse to each $CP^1$. For such a configuration the 
non-abelian gauge group is
$SU(2)^8\times (SU(2)')^8$. We identify a similar configuration in F-theory 
which has identical gauge group and for which the background $\lambda$
agrees with that of the GP model in the weak coupling limit. We then
consider various deformations away from this point in the GP model
and identify corresponding deformations in F-theory by matching a) the 
unbroken gauge group and b) the background axion-dilaton field in the
weak coupling limit. We find exact one to one correspondence between
deformations in the GP model and those in the F-theory. We also
consider special subspaces in the moduli space where the gauge group is
enhanced on the GP side, typically giving $Sp(2k)$ or $SU(2k)$ gauge groups.
We identify the corresponding symmetry enhancement points on the F-theory
side, and again find exact agreement between codimensions of these subspaces 
in the two theories.

In a previous paper\cite{PREV} we 
had analysed part of this problem by studying
the physics near one pair of intersecting orientifold planes 
accompanied by four pairs of intersecting seven branes. It was found that
non-perturbative effects in the orientifold theory converts the
background axion-dilaton field into an F-theory like configuration. 
The present paper generalizes this result in two important ways. First 
it shows how to put sixteen copies of this structure together to get
the full GP model. Second it shows how to describe in F-theory
switching on of the vacuum expectation value of the massless fields
associated with the open string states stretched between an 
intersecting pair of D-branes. We also provide an explanation for an
apparent discrepancy between the enhanced gauge symmetries in the two
theories for a configuration where the axion-dilaton field is constant
on the base.

The plan of this paper is as follows. In section \ref{s2} we give a
brief review of F-theory on the elliptically fibered Calabi-Yau manifold
over
the base $CP^1\times CP^1$, and also of the T-dual of the GP model.
In section \ref{s3} we identify the $SU(2)^8\times (SU(2)')^8$
family of points in the GP model with a specific family of points in the
moduli space of the F-theory, and also compare  deformations away from
these points in both theories. In this context we consider both,
deformations that are neutral under the $SU(2)^8\times(SU(2)')^8$
gauge group, and deformations that are charged under the gauge group, and
find exact correspondence between these deformations in the two theories. 
The axion-dilaton background in the two theories also agree in the weak
coupling limit.
Later in this section we consider the reverse problem, namely finding 
subspaces of the moduli space in both theories where the gauge symmetry is
enhanced. Again we find exact one to one correspondence 
between these subspaces in the two theories. 
In section \ref{s4} we summarize our results with some concluding
remarks.

\sectiono{Review of F-theory and Gimon-Polchinski Model} \label{s2}

We start with a review of F-theory on the Calabi-Yau manifold with
elliptic fibration over $CP^1\times CP^1$. Let $u$ and $v$ denote the
complex coordinates of the two $CP^1$'s. Also let us define the complex
scalar field $\lambda$ as
\be \label{e5}
\lambda = a + i e^{-\Phi}\, ,
\ee 
where $a$ is the axion field and $\Phi$ is the dilaton field.
Then this particular F-theory
compactification may be described as type IIB theory compactified on
$CP^1\times CP^1$, with a background $\lambda(u,v)$ equal to the complex
structure modulus of the torus described by the equation:
\be \label{e10}
y^2=x^3 + f(u,v) x + g(u,v)\, .
\ee
Here $f$ and $g$ are polynomials in $u$ and $v$ of degree (8,8) and (12,12)
respectively. The coefficients appearing in $f$ and $g$ are part of the
moduli of this F-theory compactification. From \refb{e10} we can write
down a more explicit form of $\lambda$ as function of $u$ and $v$:
\be \label{e11}
j(\lambda(u,v))= {4\cdot (24f)^3\over 4f^3 + 27 g^2}\, ,
\ee
where $j(\lambda)$ denotes the modular invariant function of $\lambda$
with a single pole at $\lambda=i\infty$, zero at $\lambda=e^{i\pi/3}$
and normalized as in ref.\cite{SENF}.
Thus at the zeroes of the denominator
\be \label{e15}
\Delta \equiv 4 f^3 + 27 g^2
\ee
$\lambda$ goes to $i\infty$ up to an SL(2,Z) transformation. These 
surfaces of (complex) codimension one\footnote{Throughout this paper we shall
count complex codimension / dimension 
of various subspaces unless specified
otherwise.} 
can be identified with the locations of the seven-branes in this background.

Generically, an F-theory compactification on a Calabi-Yau manifold with
Hodge numbers $(h_{11}, h_{12})$ has $h_{12}+1$ neutral hypermultiplets.
This gives a $2(h_{12}+1)$  (complex) dimensional hypermultiplet moduli
space. However, of these only a $h_{12}$ dimensional subspace is 
visible as complex structure deformation of the Calabi-Yau 
manifold,\footnote{This point has also been emphasized recently in 
refs.\cite{FMW,BJPS}.} and
can be identified with the coefficients appearing in the polynomials
$f$ and $g$. The rest of the deformations in the hypermultiplet moduli
space is non-geometrical and is not visible as complex structure deformations
of the Calabi-Yau manifold. We need to keep this in mind when we try to
compare the moduli space deformations in the GP model with those in F-theory.

Let us now turn to a brief description of the
T-dual version of the GP model that we shall study.
This may be described as type IIB string theory $T^2\times (T^2)'\times 
R^6/(Z_2 \times Z_2')$ where we label $T^2$ and $(T^2)'$  by $(x^6, x^7)$, 
and $(x^8, x^9)$ respectively, and $Z_2$ and $Z_2'$ are generated by
\be \label{e1}
 g=(-1)^{F_L}\cdot\Omega\cdot\II_{67},\qquad 
h=(-1)^{F_L}\cdot\Omega\cdot\II_{89}.
\ee
Here $\II_{67}$ and $\II_{89}$ denote the transformations
$(x^6\to -x^6, x^7\to -x^7)$ and $(x^8\to -x^8, x^9\to -x^9)$
respectively, $(-1)^{F_L}$ denotes the transformation that changes the 
sign of all the Ramond sector states on the left moving sector of the world
sheet of string theory, and $\Omega$ denotes the world-sheet parity
transformation.\footnote{Eq.\refb{e1} does not completely specify the
action of $g$ and $h$ on twisted sector / open sring states. These are
determined by demanding that this model is related to  the
GP model by T-duality. This was discussed at length in \cite{PREV}.} 
If we define
\be \label{e2a}
w = x^6+ ix^7, \qquad z=x^8+ix^9\, ,
\ee
and if $\tau$ and $\tau'$ denote the complex structure moduli of $T^2$ 
and $(T^2)'$ respectively, then the seven planes at
\be \label{e3a}
w = 0, {1\over 2}, {\tau\over 2}, {\tau+1\over 2}\, ,
\ee
and 
\be \label{e3b}
z = 0, {1\over 2}, {\tau'\over 2}, {\tau'+1\over 2}\, ,
\ee
are fixed under $g$ and $h$ respectively. These are known as orientifold
seven planes. We can
introduce $g$ and $h$ invariant coordinates $u$ and $v$ through the relations
\be \label{e4}
dw = du \prod_{m=1}^4 (u -\wt u_m)^{-{1\over 2}}, \qquad
dz = dv \prod_{m=1}^4 (v -\wt v_m)^{-{1\over 2}}. 
\ee
The numbers $\{\wt u_m\}$ ($\{\wt v_m\}$) are to be chosen such that 
their images in the $w$ ($z$) plane correspond to the values given in
\refb{e3a} (\refb{e3b}). This relates the parameters $\tau$ and $\tau'$ to
the cross ratios 
\be \label{e4a}
{(\wt u_1 - \wt u_2) (\wt u_3 - \wt u_4)\over (\wt u_1 - \wt u_3)
(\wt u_2 - \wt u_4)}\, \qquad 
and \qquad {(\wt v_1 - \wt v_2) (\wt v_3 - \wt v_4)\over (\wt v_1 - \wt v_3)
(\wt v_2 - \wt v_4)}\, ,
\ee
respectively. In the $(u,v)$ coordinate system    the orientifold seven
planes are located at $u=\wt u_m$ and $v=\wt v_m$ respectively.

It is known from standard analysis (see {\it e.g.} \cite{GIMPOL})
that each orientifold plane carries
$-4$ units of RR charge. In other words, along any closed contour $C$
enclosing an orientifold seven plane,
\be \label{e6}
\ointop_C d\lambda = -4\, ,
\ee
signalling that the axion $a$ changes by $-4$ units as we move around an
orientifold plane.
Since the $u$ and $v$ planes are compact  ($CP^1$) this RR charge must be
neutralized. This is done by putting 
sixteen Dirichlet seven branes transverse to
the $u$  plane and sixteen Dirichlet seven branes transverse to the $v$ 
plane\cite{POLCHI}. For reasons explained in \cite{PREV} 
these seven branes move only
in pairs. Thus a generic configuration will have eight pairs of
D-branes placed at $u=u_i$ ($1\le i\le 8$) and eight pairs of
D-branes placed at $v=v_i$ ($1\le i\le 8$). For any contour $C$ 
enclosing such a D-brane pair, we have
\be \label{e7}
\ointop_C d\lambda = 2\, .
\ee

This particular orientifold has $N=1$ supersymmetry in six dimensions.
The massless spectrum is as follows:
\begin{enumerate}
\item
The untwisted sector closed string states contribute 
\begin{enumerate}
\item The N=1 supergravity multiplet, 
\item A massless tensor multiplet,  
and 
\item Four hypermultiplets. These contain the
axion-dilaton field $\lambda$, and the moduli $\tau$ and $\tau'$.
The other moduli associated with these hypermultiplets 
will not be visible as deformations of the complex structure 
moduli of the Calabi-Yau manifold on the F-theory side.
This is a reflection of the fact that of the $2(h_{12}+1)$ 
hypermultiplet moduli in F-theory, onle $h_{12}$ are visible as complex
structure deformations of the Calabi-Yau manifold.
\end{enumerate}

\item
The closed string states twisted by the element
\be \label{e8}
gh=(-1)^{F_L+F_R}\II_{67}\II_{89}\, ,
\ee
contribute 16 hypermultiplets from the sixteen fixed points of 
$\II_{67}\II_{89}$. These contain the blow up modes of the corresponding
orbifold singularities.

\item
The open string states with ends lying on D-branes that are parallel to 
each other contribute
\begin{enumerate}
\item Massless vector multiplets corresponding to the 
gauge group $SU(2)^8\times (SU(2)')^8$. Each $SU(2)$ is associated with a
D-brane pair, with the first eight being associated with 
the pairs at $u=u_i$ and the last eight being associated with the pairs
at $v=v_i$.
\item Sixteen gauge neutral hypermultiplets, containing the locations $u_i$,
$v_i$ of the D-brane pairs. Note that as in the case of F-theory, only
half of each hypermultiplet represents geometrical modulus.
\end{enumerate}

\item
Finally open string states whose ends lie on intersecting D-branes
contribute massless hypermultiplets in the (2,2) representation
of $SU(2)_i\times SU(2)'_j$ for all pairs $(i,j)$ ($1\le i,j\le 8$).
Here $SU(2)_i$ denotes the $SU(2)$ group associated with the D-brane pair
at $u=u_i$ and $SU(2)'_j$ denotes the $SU(2)$ group associated with the 
D-brane pair at $v=v_j$.
We shall denote these hypermultiplets as $(2_i,2'_j)$ states.
\end{enumerate}

In this model we can further break the $SU(2)^8\times (SU(2)')^8$ group
by giving vev to the $(2_i,2'_j)$ hypermultiplets. 
We shall now analyze some specific breaking patterns which will be useful for
later study. Consider giving vev to the $(2_i,2'_j)$ states
for all $(i,j)$ with $1\le i\le p$, $1\le j\le q$. This leaves
$SU(2)^{8-p}\times (SU(2)')^{8-q}$ unbroken. Let $SU(2)_d$ denotes the
diagonal subgroup of the first $p$ $SU(2)$ and the first $q$ $SU(2)'$
groups. Then under $SU(2)_d$ the $(2_i,2'_j)$ state decomposes as
${\bf 3 \oplus 1}$. Two of the possible symmetry breaking patterns are the
following:
\begin{enumerate} 
\item We can consider giving vev to only the singlet components
of all $(2_i,2'_j)$ states. This breaks the first $p$ $SU(2)$ and the
first $q$ $SU(2)'$ to $SU(2)_d$ in general. The resulting massless spectrum
consists of 
\be \label{eaa}
pq
\ee
neutral hypermultiplets, and
\be \label{ebb}
(p-1)(q-1)
\ee
hypermultiplets in the triplet of $SU(2)_d$, taking into account the fact
that $(p+q-1)$ of the triplet 
hypermultiplets become massive by Higgs mechanism.

\item 
For $p>1$, $q>1$, we can further break $SU(2)_d$ by giving vev to the 
surviving massless triplets
of $SU(2)_d$. For $p=q=2$ we have only one triplet of $SU(2)_d$ and hence 
vev of this triplet breaks $SU(2)_d$ to U(1) leaving one extra neutral 
hypermultiplet. On the other hand for $p\ge 3$ or $q\ge 3$ we have more than
one triplet of $SU(2)_d$ and giving vev to these triplets we can break
$SU(2)_d$ completely. The number of extra neutral hypermultiplets,
obtained after this symmetry breaking, is given by
\be \label{ecc}
3(p-1)(q-1)-3 = 3(pq-p-q)\, .
\ee

\end{enumerate}

In particular, by taking $p=q=8$, and carrying out the second step in the
above description, we can break the gauge group completely.

Instead of breaking the $SU(2)^8\times (SU(2)')^8$ group, we can also
enhance the gauge group further by bringing the D-brane pairs on top 
of each other, and/or on top of an orientifold plane. If $k$ D-brane
pairs are on top of each other the gauge group is $Sp(2k)$, whereas if
$k$ D-brane pairs are on top of an orientifold plane, the
non-abelian part of the gauge group is $SU(2k)$.

\sectiono{Comparison of the Two Theories} \label{s3}

In this section we shall carry out a detailed comparison between the
two theories discussed in the previous section. Naively one
would have thought that a convenient starting point would be a
configuration where the RR charge is neutralized locally in the
internal space, and hence the axion-dilaton field is a constant,
as was the case in \cite{SENF}. 
In the GP model this would correspond to placing
two D-brane pairs on top of each orientifold plane, thereby giving
an $(SU(4))^8$ non-abelian gauge group. On the other hand, in F-theory,
this would correspond to choosing 
\be \label{eqn1}
f(u,v) = \alpha \prod_{m=1}^4 (u - \wt u_m)^2 (v - \wt v_m)^2, \qquad
g(u,v) = \prod_{m=1}^4 (u - \wt u_m)^3 (v - \wt v_m)^3\, ,
\ee
where $\alpha$ is an arbitrary constant. This corresponds to a set of
intersecting $D_4$ singularities giving rise to 
$SO(8)^8$ gauge group as well as
tensionless strings\cite{VAFSIX,BERJOH}. 
Thus the two theories would seem to disagree.
It turns out that there is a subtle reason for this discrepancy that will
be explained later in this section. However, due to this subtlety,
this will not  be a convenient starting point for our analysis.

Instead we shall start from the point in the moduli space of the GP model
with $SU(2)^8\times (SU(2)')^8$  gauge symmetry 
and identify the corresponding
point in the F-theory moduli space. We shall then consider various symmetry
breaking as well as symmetry enhancement patterns as we move
in the moduli spaces of
the two theories and compare the results.

\subsection{The $SU(2)^8\times (SU(2)')^8$ Point}

In the GP model this corresponds to a configuration of sixteen pairs
of D-branes situated at  $u=u_i$ and $v=v_i$ ($1\le i\le 8$)
and eight orientifold planes situated at $u=\wt u_m$ and $v=\wt v_m$
($1\le m\le 4$). Requiring $\lambda$ to be an analytic function of
$u$ and $v$ (for preservation of space-time supersymmetry) and 
using eqs.\refb{e6} and \refb{e7} we get the following behaviour of
$\lambda$ near the D-branes and the orientifold planes:
\ben \label{e12}
\lambda &\simeq& {2\over 2\pi i} \ln (u-u_i) \quad \hbox{as} \quad u\to u_i,
\nonumber \\
\lambda &\simeq& {2\over 2\pi i} \ln (v-v_i) \quad \hbox{as} \quad v\to v_i,
\een
\ben \label{e14}
\lambda &\simeq& -{4\over 2\pi i} \ln (u-\wt u_m) 
\quad \hbox{as} \quad u\to \wt u_m,
\nonumber \\
\lambda &\simeq& -{4\over 2\pi i} \ln (v-\wt v_m) 
\quad \hbox{as} \quad v\to \wt v_m.
\een
{}From \refb{e12} we 
see that as $u\to u_i$ or $v\to v_i$, $\lambda$ approaches
$i\infty$. This corresponds to weak coupling and hence this 
behaviour is not expected to be
modified due to quantum corrections (except 
possible corrections to the locations
of the D-branes). If we continue to denote by $u_i$ and $v_i$ the quantum
corrected locations of the D-branes, then we expect the following 
behaviour of $j(\lambda)$ near $u=u_i$ and $v=v_i$:
\ben \label{e13}
j(\lambda) &\simeq & {1\over (u-u_i)^2} \quad \hbox{as} \quad u\to u_i\, ,
\nonumber \\
&\simeq & {1\over (v-v_i)^2} \quad \hbox{as} \quad v\to v_i\, .
\een
On the other hand as $u\to \wt u_m$ or $v\to \wt v_m$, $\lambda$
approaches $-i\infty$. This is inconsistent with the definition \refb{e5} of
$\lambda$   according to which the imaginary part of $\lambda$ is
positive definite. Thus strong coupling effects must modify
this behaviour. From our analysis of ref.\cite{SENF} we know what kind of
modification we should expect. Away from the intersection
points, each orientifold plane should split into two
seven branes related to the D-brane by SL(2,Z) transformation, such that
the total monodromy of $\lambda$ as we go around both these branes is
given by \refb{e6}. In the weak coupling limit the splitting between these
two branes is small. Thus after taking into account these 
quantum corrections,
we should see a pair of poles of $j(\lambda)$ around each
of the surfaces $u=\wt u_m$ and $v=\wt v_m$.

In order to reproduce this behaviour in F-theory, we need to adjust the
functions $f$ and $g$ such that $j(\lambda)$ calculated from eq.\refb{e11}
has these properties. First of all, in order to reproduce \refb{e13},
$\Delta$ defined in
eq.\refb{e15} should be of the form:
\be \label{e16}
\Delta = \Big( \prod_{i=1}^8 (u-u_i)^2 (v-v_i)^2 \Big) \delta(u,v),
\ee
where $\delta$ is a polynomial of degree (8,8).
At the first sight it would seem extremely unlikely 
that it will be possible to
find such $f$ and $g$.  Since $f$ and $g$ are polynomials of
degree (8,8) and (12,12) respectively, the total number of
adjustable parameters is given by
\be \label{e17}
9^2 + (13)^2 = 250\, .
\ee
$\Delta$ is a polynomial of degree (24,24) in $f$ and $g$. 
Thus demanding that
$\Delta$ has the form \refb{e16} with $u_i$ and $v_i$ 
fixed according to our choice gives 
\be \label{e18}
(25)^2 - 9^2 = 544
\ee
constraints on the coefficients appearing in $f$ and $g$. Here $9^2$ 
reflects the number of
coefficients appearing in $\delta$ which can be adjusted in  trying to
solve eq.\refb{e16}. This is clearly a highly overdetermined
system of equations.

Nevertheless eq.\refb{e16} has a simple solution. Let us
choose:
\be \label{e18a}
f = \eta -3h^2 \, ,
\ee
and
\be \label{e19}
g = h ( \eta - 2h^2)\, ,
\ee
where $h$ is a polynomial of degree (4,4) and $\eta$ 
is a polynomial of degree
(8,8)  in $u$ and $v$ respectively. In this case, $\Delta$ calculated
from eq.\refb{e15} is given by
\be \label{e20}
\Delta = (4\eta - 9h^2) \eta^2\, .
\ee
Thus we can now easily satisfy the requirement \refb{e16} by
choosing
\be \label{e21}
\eta(u,v) = C \prod_{i=1}^8 (u-u_i) (v-v_i)\, ,
\ee
where $C$ is an arbitrary constant. This gives,
\be \label{e22}
\Delta(u,v) = C^2 \prod_{i=1}^8 (u-u_i)^2 (v-v_i)^2
\big( 4C \prod_{i=1}^8 (u-u_i) (v-v_i) - 9 h^2\big)\, ,
\ee
and 
\be \label{e23}
j(\lambda) =
{4\cdot (24)^3\cdot  
\big( C \prod_{i=1}^8 (u-u_i) (v-v_i) - 3 h^2\big)^3 \over
C^2 \prod_{i=1}^8 (u-u_i)^2 (v-v_i)^2
\big( 4C \prod_{i=1}^8 (u-u_i) (v-v_i) - 9 h^2\big)}\, .
\ee
{}From this expression we see that as $C\to 0$,
with $\{u_i\}$, $\{v_i\}$ and $h(u,v)$ fixed, $j(\lambda)\to \infty$
almost everywhere in the $(u,v)$ space except  on subspaces of
codimension one where the numerator vanishes. Thus $C\to 0$ corresponds
to the weak coupling limit of the theory.  In this limit, we can associate
the deformations associated with $C$ to the deformations in the
GP model associated with the vev of $\lambda$.

In order to correctly
reproduce the configuration in the GP model, we must also
examine the other zeroes of $\Delta$ and ensure that they lie pairwise
near the surfaces $u=\wt u_m$ and $v=\wt v_m$ for small $C$. From \refb{e22}
we indeed observe that for small $C$ the 
other zeroes of $\Delta$ lie pairwise around the surface 
\be \label{e24}
h(u,v) = 0\, .
\ee
Thus we must adjust $h$ such that the surface $h=0$ 
coincides with the locations
of the orientifold planes. This is easily done by choosing:
\be \label{e25}
h(u,v) = K \prod_{m=1}^4 (u-\wt u_m) (v - \wt v_m)\, ,
\ee
where $K$ is an arbitrary constant. There is however some redundancy
in this parametrization of $h$. First of all, note that $j(\lambda)$
defined in \refb{e11} is invariant under a rescaling of the form:
\be \label{e26}
f\to s^2 f, \qquad g\to s^3 g\, ,
\ee
for any constant $s$.
Using this freedom we can choose $K$ to be unity. Furthermore, since both 
$u$ and $v$ parametrize $CP^1$, there is a pair of
SL(2,C) transformations on $u$ and $v$ which simply reflect different choice
of coordinates on the base. 
Using these transformations we can fix three of the 
$\wt u_m$ and three of the $\wt v_m$ to any value we like. Thus at the end the
relevant parameters appearing in $h$ are the cross ratios defined in
eq.\refb{e4a}. As has already been pointed out, these cross ratios correspond
to the moduli $\tau$ and $\tau'$ in the GP model.

It is however clear that in F-theory, if we want to get $SU(2)^8\times 
(SU(2)')^8$ gauge group, then all we need is sixteen $A_1$ singularities
where $\Delta$ has double zeroes without $f$ and $g$ vanishing. This
means that we maintain the $SU(2)\times (SU(2)')^8$ symmetry even
when we choose $h$ to be completely arbitrary, since according to
eq.\refb{e22} this does not affect the double zeroes of $\Delta$.
If we are to establish a one to one correspondence between the deformations
in F-theory and those in the GP model, then we must 
be able to interprete the deformations
associated with $h$ as some deformations in the GP model. For this it will 
be convenient to choose a
specific parametrization of $h$. $h$, to begin with, has $5^2=25$ 
coefficients, but the freedom of rescaling $f$ and $g$ and SL(2,C)
transformations on $u$ and $v$ remove seven of these parameters. Two
of the remaining eighteen parameters are already present in the form
of $h$ given in \refb{e25} in the cross ratios \refb{e4a}. Thus we 
need to introduce sixteen more parameters in $h$. We choose the following
parametrization:
\be \label{e29}
h(u,v) = \prod_{m=1}^4(u-\wt u_m)(v-\wt v_m) +\sum_{k,l=1}^4
\alpha_{kl} \prod_{m\ne k} (u-\wt u_m)\prod_{n\ne l} (v -\wt v_m)\, ,
\ee
with $\wt u_1$, $\wt u_2$, $\wt u_3$, $\wt v_1$, $\wt v_2$ and $\wt v_3$ 
chosen to be some fixed numbers. $\alpha_{kl}$ are the new parameters 
introduced for labelling the most general $h$ up to the redundancy
discussed above.

We shall now show that the deformations associated with $\alpha_{kl}$
correspond in GP model to the effect of blowing up the sixteen 
orbifold singularities by switching on 
the twisted sector closed string states.
For this let us consider the deformation where 
$\alpha_{11}\ne 0$, and all other
$\alpha_{ij}$ are zero. Then we have
\be \label{e30}
h(u,v)= \prod_{m=2}^4 (u-\wt u_m) (v - \wt v_m) \{ (u-\wt u_1)
(v - \wt v_1) + \alpha_{11}\}\, .
\ee
As we have already seen from eq.\refb{e22}, in the weak coupling limit
($C\to 0$) the approximate locations of the orientifold
planes are at the zeroes of $h$. Thus the orientifold planes are now
situated at:
\ben \label{e31}
u & = & \wt u_m, \quad \hbox{for} \quad 2\le m\le 4\, ,\nonumber \\
v & = & \wt v_m, \quad \hbox{for} \quad 2\le m\le 4\, ,\nonumber \\
\hbox{and} && (u-\wt u_1) (v - \wt v_1) + \alpha_{11} =0\, .
\een
In other words, the orientifold planes at $u=\wt u_1$ and $v=\wt v_1$ have
joined together to become a smooth complex hyperbola.

Thus we now need to show that this can be interpreted as the 
result of blowing up the orbifold singularity at the 
intersection of $u=\wt u_1$ and $v=\wt v_1$.
For this it will be useful to take the generators of the
$Z_2\times Z_2'$ group as:
\be \label{e32}
gh = \II_{67}\II_{89} (-1)^F, \qquad \hbox{and} \qquad g = 
(-1)^{F_L}\cdot\Omega \cdot \II_{67}\, .
\ee
We shall first mod out the type IIB theory on $T^4$ by the $Z_2$
group generated by $gh$, then blow up the orbifold singularity at 
$u=\wt u_1$, $v=\wt v_1$,
and finally mod out the theory by the $Z_2$ group generated by $g$ and
study the location of the orientifold planes.
Let the image of the point $(\wt u_1, \wt v_1)$ in the $(w,z)$ plane
be at $(w=0,z=0)$
with $z$ and $w$ as defined in eq.\refb{e2a}.
Since $gh$ takes $z$ to $-z$ and $w$ to $-w$, close to $(w=0,z=0)$
the
single valued coordinates on the resulting space are:
\be \label{e33}
u-\wt u_1 = w^2, \qquad v-\wt v_1=z^2, \qquad \xi= zw,
\ee
with the restriction
\be \label{e34}
(u-\wt u_1)(v-\wt v_1) = \xi^2\, .
\ee 
The surface described by this equation is 
singular at $(u=\wt u_1, v=\wt v_1)$
reflecting the orbifold singularity present there. We can blow up the
singularity by replacing eq.\refb{e34} by
\be \label{e36}
(u-\wt u_1)(v-\wt v_1) - \xi^2 = a^2\, ,
\ee 
where $a$ is the blow up parameter. 

Now we mod out type IIB theory on this blown up space by the $Z_2$
group generated by $g$. Since $g$
changes the sign of $w$ without changing the sign of $z$, we see from 
eq.\refb{e33} that under this transformation $u$ and $v$ remain unchanged
but $\xi$ changes sign. Thus the orientifold plane is situated at
$\xi=0$, which using eq.\refb{e36} can be rewritten as
\be \label{e37}
(u-\wt u_1) (v-\wt v_1) = a^2\, .
\ee
Comparing with \refb{e31} we see that this is precisely what we got on the
F-theory side provided we identify $\alpha_{11}$ with $-a^2$. Thus 
the deformation associated with $\alpha_{11}$ indeed represents the blow
up mode in the GP model of the orbifold singularity at $(\wt u_1,
\wt v_1)$. Similar analysis shows that $\alpha_{kl}$ represents the 
blow up mode of the orbifold singularity at $(\wt u_k, \wt v_l)$.

This exhausts all deformations that are neutral under 
$SU(2)^8\times (SU(2)')^8$. We shall now turn to the effect of
switching on the vev of the hypermultiplets which are charged under
this gauge group.

\subsection{Symmetry Breaking Pattern}

We now turn to the deformations in the GP model due to switching on the
vev of the $(2_i, 2'_j)$ states, and identify the corresponding deformations
on the F-theory side. (Similar analysis has been carried out in 
ref.\cite{VAFSIX}
in the context of intersecting singularities in F-theory.)
In particular we shall consider a specific symmetry breaking pattern in 
which we switch on the vev of all the $(2_i, 2'_j)$ states for
$(1\le i\le p)$, $(1\le j\le q)$. As pointed out in the last section, by
aligning these vev's properly one can break the first $p$ $SU(2)$
and the first $q$ $SU(2)'$ into their diagonal subgroup 
$SU(2)_d$. The last $(8-p)$ $SU(2)$ and the last $(8-q)$ $SU(2)'$ remain
unbroken. We shall first try to understand this subspace in F-theory. At
the end we shall also study the case where we destroy the alignment of the
vev's and break $SU(2)_d$ as well.

First of all, since $SU(2)^{8-p}\times (SU(2)')^{8-q}$ is preserved 
$\Delta$ should still possess a factor
\be \label{e38}
\prod_{i=p+1}^8 (u-u_i)^2 \prod_{j=q+1}^8 (v-v_i)^2\, .
\ee
Second, since $SU(2)_d$ is preserved, one would expect that 
the deformation on the F-theory side should convert the 
$$ \prod_{i=1}^p (u-u_i)^2 \prod_{j=1}^q (v-v_i)^2 $$
factor in \refb{e22} into a perfect square:
\be \label{e39}
(\phi_{p,q}(u,v))^2\, ,
\ee
where $\phi_{p,q}$ is a polynomial of degree $(p,q)$ in $(u,v)$. This
would ensure the presence of an $A_1$ singularity and hence unbroken
$SU(2)_d$. From eqs.\refb{e18a}-\refb{e20} we see that this form of
$\Delta$ can easily be achieved if we choose $f$ and $g$ of the form
\refb{e18a} and \refb{e19} respectively, with arbitrary $h$ and
\be \label{e40}
\eta(u,v) = C \phi_{p,q}(u,v)
\prod_{i=p+1}^8 (u-u_i) \prod_{j=q+1}^8 (v - v_i)
\, .
\ee
We shall now compare the number of extra deformation parameters that
appear in the two theories.\footnote{Note that the agreement between
the numbers of neutral hypermultiplets
in the two theories would be a trivial consequence
of the anomaly cancellation condition in six dimensions if we could 
establish that these two theories have the same gauge group and same
charged matter content. But in
F-theory we cannot easily identify U(1) factors
in the gauge group. There is also some ambiguity in determining the
spectrum of charged matter in F-theory although much progress in this
direction has been made\cite{KATVAF}.}
In GP model it is given by \refb{eaa}.
On the F-theory side the extra deformation parameters are those
appearing in $\phi_{p,q}$, modulo the parameters that were already
present earlier. Since $\phi_{p,q}$ is a polynomial of degree 
$(p,q)$, it has $(p+1)(q+1)$ parameters to start with. Of this the
coefficient of $u^pv^q$ can be absorbed into the parameter $C$.
Furthermore $(p+q)$ of these parameters were present before in the form
of $u_i$ ($1\le i\le p$) and $v_j$ ($1\le j\le q$). Thus the net number
of extra parameters is
\be \label{e41}
(p+1)(q+1) - 1 - p -q =pq\, ,
\ee
in perfect agreement with the answer \refb{eaa} in the GP model.

For the special case $p=q=1$ we can take
\be \label{e42}
\phi_{1,1}(u,v) = (u-u_1)(v-v_1)+\beta\, ,
\ee
where $\beta$ is an arbitrary constant.
$\phi_{1,1}=0$
is the location of the D-brane after switching on the vev of the
($2,2'$) state. Thus we see that the effect of this deformation is to
fuse two D-branes into one smooth complex hyperbola (times a five dimensional
manifold).

Finally in the GP model
we consider deformations that break $SU(2)_d$ as well by switching 
on vev of the hypermultiplets in the triplet of $SU(2)_d$.
The number of such triplets is $(p-1)(q-1)$ as
given in eq.\refb{ebb} and hence
this breaking is possible only for $p>1$ and $q>1$. For $p=q=2$ $SU(2)_d$
is broken to $U(1)$, and we get an extra neutral hypermultiplet after
the breaking. For $p\ge 3$ or $q\ge 3$, $SU(2)_d$ can be broken
completely, and we get $3(pq-p-q)$ extra neutral hypermultiplets after
this breaking, as given in eq.\refb{ecc}.
On the F-theory side breaking of $SU(2)_d$
would mean that we no longer require $\Delta$
to contain a factor of the form $\phi_{p,q}^2$. Thus we can now relax
\refb{e18a}, \refb{e19}.
However, we would still want $\Delta$ to contain a factor of
\be \label{eeta}
\eta_0^2\equiv \prod_{i=p+1}^8 (u-u_i)^2\prod_{j=q+1}^8 (v-v_i)^2 
\ee
since $SU(2)^{8-p}\times (SU(2)')^{8-q}$ is still unbroken. 
For this let us consider the following deformations of 
eqs.\refb{e18a}, \refb{e19}
\ben \label{e43}
f &=& \eta - 3h^2 + \delta f \nonumber \\
g &=& h(\eta-2h^2) + \delta g\, ,
\een
where $h$ is an arbitrary polynomial of degree (4,4) in ($u,v$) and 
$\eta$, according to eqs.\refb{e40}, \refb{eeta} can be expressed as
\be \label{eeta2}
\eta = C \eta_0 \phi_{p,q}\, .
\ee
We shall work to first order in $\delta f$ and $\delta g$ and find 
the number of independent deformations for which
\be \label{emm8}
\delta \Delta = 12 f^2 \delta f + 54 g \delta g\, ,
\ee
has a factor of $\eta_0^2$.
Using eqs.\refb{e43} we can reexpress \refb{emm8} as
\be \label{emm9}
\delta \Delta = 108 h^3 (h \delta f - \delta g) + 18 h \eta_0\phi_{p,q}
(3 \delta g - 4h \delta f) + 12 \eta^2_0\phi_{p,q}^2 \delta f\, .
\ee
The last term already has a factor of $\eta_0^2$ and hence can be 
ignored during the rest of the analysis. Since the second term has an
explicit factor of $\eta_0$, a necessary condition for
the full expression to have a factor of $\eta_0^2$ is 
that the first term must also have a factor of $\eta_0$.
Thus we have the requirement:
\be \label{emm10}
\delta g - h\delta f = \eta_0\delta\chi_{p+4, q+4}\, ,
\ee
where $\delta\chi_{p+4,q+4}$ is a polynomial of degree $(p+4,q+4)$ 
in $(u,v)$ since $\eta_0$ is of degree $(8-p,8-q)$. 
Substituting this in \refb{emm9} and ignoring terms of order
$\eta_0^2$ we get
\be \label{emm11}
\delta \Delta = -18 h^2 \eta_0 (\phi_{p,q}\delta f 
+ 6h \delta\chi_{p+4,q+4})\, .
\ee
Thus in order that $\delta\Delta$ is proportional to 
$\eta_0^2$, we require that
the expression inside (~) be proportional to $\eta_0$:
\be \label{emm12}
\phi_{p,q}\delta f + 6 h \delta\chi_{p+4,q+4} = \eta_0 \delta\xi_{2p,2q}\, ,
\ee
for some polynomial $\delta\xi_{2p,2q}$ of degree $(2p,2q)$ in $(u,v)$.

Thus the question to be addressed is: how many independent solutions of
the above equation do we have? Since $\delta f$, $\delta\chi_{p+4,q+4}$
and $\delta\xi_{2p,2q}$ are of degree (8,8), $(p+4, q+4)$ and $(2p, 2q)$
in $(u,v)$ respectively, the total number of adjustable parameters is
\be \label{emm13}
N_{par} = 81 + (p+5)(q+5) + (2p+1)(2q+1)\, .
\ee
Both sides of eq.\refb{emm12} are polynomials of degree $(8+p, 8+q)$.
Thus the total number of constraints is
\be \label{emm14}
N_{con} = (p+9)(q+9)\, .
\ee
Naively, we would have $N_{par}-N_{con}$ solutions of eq.\refb{emm12}.
However, one should keep in mind that some of these deformations may
simply correspond to deformations of $h$, $C$ and $\phi_{p,q}$, which 
move us inside the subspace of unbroken $SU(2)_d$, and some may simply
correspond to the redundant deformations associated with the SL(2,C)
transformations on $u$ and $v$ or rescaling of $f$ and $g$ given in 
\refb{e26}. All such deformations can be described by deforming 
$h$ by an arbitrary polynomial of degree (4,4) and deforming $C\phi_{p,q}$
by an arbitrary polynomial of degree $(p,q)$. Thus the total number of
redundant deformations is given by
\be \label{emm15}
N_{red} = 25 + (p+1) ( q+1)\, .
\ee
This gives the total number of independent deformations that take us out
of the subspace of unbroken $SU(2)_d$ as:
\be \label{emm16}
N_{par}-N_{con}-N_{red} = 3 (pq - p -q)\, ,
\ee
in exact agreement with the GP model answer \refb{ecc}.

Note however that for $p=q=2$ \refb{emm16} vanishes, whereas in the GP
model this corresponds to a one parameter family of deformations.
This seems to be a contradiction, but this is resolved
by the fact that in this case all the $N_{con}$ constraints are not
independent. Indeed since now $\eta_0$ is of degree (6,6), 
one can explicitly construct a one parameter family
of solutions for $\delta f$ and $\delta g$ satisfying all the
requirements as follows:
\be \label{emm17}
\delta f = 0\, , \qquad \delta g = \gamma \eta_0^2\, ,
\ee
where $\gamma$ is an arbitrary constant. Thus we see that the
F-theory results are again consistent with the results from the GP model.

\subsection{Symmetry Enhancement}

We shall now start from the generic $SU(2)^8\times (SU(2)')^8$ configuration
in the two theories, and consider special subspaces of this moduli space
where there is enhanced gauge symmetry. In the GP model the starting point
is a configuration of sixteen intersecting pairs of D-branes at arbitrary
locations $u=u_i$, and $v=v_i$ ($1\le i\le 8$) with all the sixteen orbifold
singularities blown up. In F-theory we start from $f$ and $g$ of the form
given in  \refb{e18a} and \refb{e19}, with arbitrary $h$, but $\eta$
restricted to be of the form \refb{e21}.

Now in the GP model two of the $SU(2)$ groups can combine to give an 
enhanced $Sp(4)$ gauge group if the locations of two of the D-brane pairs
(say $u_1$ and $u_2$) coincide. This is a subspace of (complex) codimension
one. There also appears a hypermultiplet in the {\bf 5} representation of
$Sp(4)$. Breaking of $Sp(4)$ to $SU(2)^2$, which corresponds to separating
the two D-brane pairs, is achieved by giving vev to this hypermultiplet.
Four of the five components become massive due to Higgs mechanism, and the
remaining component gives the extra singlet of $SU(2)\times SU(2)$ that
measures the separation between the D-brane pairs.

In F-theory this process also has a simple (and obvious) description. It
simply corresponds to the codimension one subspace of the moduli space
where we choose $u_1=u_2$ in the expression \refb{e21}
for $\eta$. As a result,
\be \label{e54}
\Delta\propto (u - u_1)^4\, .
\ee
Furthermore it can easily be seen that neither $f$ nor $g$ vanish there.
Thus this corresponds to an $A_3$ singularity. Naively one might have
thought that this signals the appearance of an $SU(4)$ gauge group, but
it can be seen that in the language 
of ref.\cite{VAFSIX} (see also \cite{ASPIN}) this singularity is
generically non-split and hence $SU(4)$ is broken to $Sp(4)$ due
to the monodromy in the $u=u_1$ plane.
To see this note that we have the following expansion of $f$ and
$g$ in a power series in $(u-u_1)\equiv \wt u$:
\ben
f &=& -3 h_1^2(v) - 3 h_1(v)h_2(v)\wt u - 3 h_3(v)\wt u^2 - h_4(v)\wt u^3
+ O(\wt u^4)\cr && \nonumber \\
g &=& 2 h_1^3(v) + 3 h_1^2(v) h_2(v) \wt u + 3 (h_3(v)+{1\over 4} h_2^2(v))
h_1(v)\wt u^2 \nonumber \\
&& + \Big[ \big( {3\over 2} h_3(v) -{1\over 8} h_2^2(v)\big)
h_2(v) + h_1(v) h_4(v)\Big] \wt u^3 + O(\wt u^4)\nonumber \\
\een
where,
\ben
h_1(v) &=& h (u_1, v)\nonumber \\
h_2(v) &=& 2 \p_{u_1}h(u_1,v) \nonumber \\
h_3(v) &=& -{1\over 3} C \prod_{i=3}^8 (u_1-u_i)\prod_{j=1}^8 (v-v_j)
+{1\over 2} \p_{u_1}^2 h^2(u_1,v) \nonumber \\
h_4(v) &=& -C \prod_{i=3}^8 (u_1 - u_i) \prod_{j=1}^8 (v-v_j)
\sum_{k=3}^8 {1\over u_1 - u_k} +{1\over 2} \p_{u_1}^3 h^2(u_1,v)\, .
\een
In order that the singularity is split we need $h_1(v)$ to be a perfect
square. Since this is not so in general, we get a non-split singularity
and hence an enhanced $Sp(4)$ gauge group.

Let us now proceed further. In the GP model if we place these two 
pairs of D-branes on top of an orientifold plane, we expect to get an
enhanced $SU(4)$ non-abelian gauge group. Naively,
this is a subspace of complex codimension 
one in the previous subspace with enhanced $Sp(4)$ gauge symmetry, since
we need to adjust only one parameter $u_1$ to be equal to the location
$\wt u_1$ (say) of the orientifold plane. But as was 
shown in ref.\cite{WITSIX},
due to one loop anomaly effects, in this configuration one of the 
sixteen blow up modes becomes massive. Hence this is really a subspace of
codimension two in the previous subspace. In the language of Higgs mechanism
this phenomenon is explained as follows. There are two hypermultiplets in
the {\bf 6} representation of $SU(4)$. Giving vev to these hypermultiplets
with appropriate alignment we can break $SU(4)$ to $Sp(4)$. Each {\bf 6} 
representation of $SU(4)$ decomposes into a {\bf 5} and a singlet of $Sp(4)$.
One of the {\bf 5}'s become massive due to Higgs mechanism, and we are 
left with two extra singlets and a {\bf 5} of $Sp(4)$.

What is the corresponding subspace of the moduli space of F-theory with
enhanced $SU(4)$ gauge symmetry? It is clear that this would correspond
to the subspace where the $A_3$ singularity is split. As has been stated
before, this requires $h_1(v)\equiv h(u_1,v)$ to be a 
perfect square. Since $h_1(v)$ is
a polynomial of degree four in $v$, requiring it to be a perfect square
gives two constraints on the parameters. Thus we see that in the F-theory
the subspace of the moduli space where $Sp(4)$ gets enhanced to $SU(4)$ is
of codimension two, again in agreement with the answer in the GP model.

Let us now consider in F-theory a subspace of this moduli
space where $h(u_1,v)$ vanishes. This new subspace is of codimension
three in the previous subspace. In this case,
\be \label{en1}
h(u,v) = (u-u_1) P_{3,4} (u,v)\, ,
\ee
where $P_{3,4}(u,v)$ is a polynomial of degree (3,4) in $(u,v)$. This
corresponds to choosing in \refb{e21}, \refb{e29}:
\ben \label{en2}
&& \wt u_1 = u_1 = u_2\nonumber \\
&& \alpha_{1l}= 0 \, \qquad \hbox{for} \qquad 1\le l\le 4\, .
\een
This gives 
\ben \label{en3}
f(u,v) &\simeq& (u-u_1)^2\, ,\nonumber \\
g(u,v) &\simeq& (u-u_1)^3\, ,
\een
near $u=u_1$. This is a $D_4$ type singularity.
In the language of ref.\cite{VAFSIX} this singularity can be
shown to be semi-split, therby giving rise to an $SO(7)$ gauge group.
Physically this can be understood as follows. Since the D-branes which
intersect the plane $u=u_1$ move only in pairs, the SL(2,Z) monodromy
in the $u=u_1$ plane around these D-brane pairs
is given by $T^2$ (with $T$
and $S$ denoting the standard generators of SL(2,Z)). According to
ref.\cite{SEIWIT} this does not induce any triality action 
in $SO(8)$. On the
other hand the monodromies around the zeroes of $\Delta$ representing the
split orientifold plane are given 
by $STS^{-1}$ and $-T^{-4}ST^{-1}S^{-1}$\cite{SEIWIT}
both of which induce $SO(8)$ triality action\cite{SEIWIT} 
\be \label{ekk1}
{\bf 8_c\leftrightarrow 8_v, \qquad 8_s\to 8_s}
\ee
where ${\bf 8_v}$, ${\bf 8_s}$ and ${\bf 8_c}$ represent the vector, 
spinor and the conjugate
spinor representations of $SO(8)$ respectively. Now there is a (non-standard)
embedding of $SO(7)$ in $SO(8)$
under which these different representations of
$SO(8)$ decompose as:
\be \label{ekk2}
{\bf 8_v = 8, \qquad 8_c=8, \qquad 8_s = 7+1\, ,}
\ee
where {\bf 7} denotes the vector representation and
{\bf 8} denotes the unique spinor representation of $SO(7)$.
{}From eq.\refb{ekk1} and \refb{ekk2} we see that the monodromy in the
$u=u_1$ plane acts trivially on this $SO(7)$ subgroup of $SO(8)$.
Thus $SO(8)$ is broken to $SO(7)$.

By requiring that the breaking of $SO(7)$ to $SU(4)$ be describable by 
conventional Higgs mechanism, we can also infer the spectrum of massless
charged
hypermultiplets in this special subspace of the F-theory moduli space.
In particular, it must contain three hypermultiplets in the 
{\bf 7} representation.
(This is consistent with the counting described in ref.\cite{SADOV}.)
Vacuum expectation values of these hypermultiplets, when 
appropriately aligned, can break $SO(7)$
to $SU(4)\equiv SO(6)$. This gives 
two hypermultiplets in the 6 representation of $SU(4)$, and
three neutral hypermultiplets,
thereby showing that the $SO(7)$ symmetry enhancement takes place inside
a codimension three subspace.

Now that we have found this subspace of enhanced $SO(7)$ symmetry in the
F-theory, we would like to ask, how do we reach these enhanced symmetry
points in the GP model? In order to analyze this question, we examine
closely the F-theory background, and try to identify the corresponding
configuration in the GP model. First, setting $u_1=u_2=\wt u_1$ implies
that in the GP model we must have two pairs of D-branes on top of a single
orientifold plane. Second, setting $\alpha_{1l}$ to zero means that we 
set to zero all the blow up modes associated with the orbifold singularities
lying on the $u=\wt u_1$ plane. However, examining the corresponding 
configuration in the GP model we still find that the gauge group is
$SU(4)$ and not $SO(7)$!

In order to understand the source of this discrepancy, let us note that 
according to our previous analysis, the three blow up modes, which had to
be switched off in order to recover the unbroken $SO(7)$ symmetry, are each
part of a vector representation of $SO(7)$. In other words there are three
hypermultiplets in the vector representation of $SO(7)$, and the components
of these which are singlet under $SU(4)\equiv SO(6)$ are the three blow
up modes. Let us recall however, that each hypermultiplet contains two
complex scalars. In the present example, the partners of a blow up mode
are the flux of two tensor fields $B_{\mu\nu}$ and $B'_{\mu\nu}$ through
the two cycle that is being blown up.\footnote{Both the tensor fields, as
well as the two cycle, are odd under the $Z_2$ generator 
$g$, and as a result
the fluxes are even.} Being super-partners of the blow up mode, these are 
also part of the vector multiplet of $SO(7)$ and hence must be switched off
in order to recover the unbroken $SO(7)$ symmetry. However, as was pointed
out by Aspinwall\cite{ASPINK3}, 
the conformal field theory orbifold has half unit
of $B$ (and possibly $B'$) flux through the two cycle switched on. This
would break $SO(7)$ to $SO(6)\equiv SU(4)$, and  thus it
is not surprising that in the GP model we do not see the $SO(7)$ gauge 
symmetry restored even when all the blow up modes are switched off.

This analysis shows that we can recover the $SO(7)$ 
gauge symmetry in the GP model
by continuously turning off the tensor field flux, but this
configuration will not, in general, be describable 
in terms of a solvable conformal field theory.
This result also shows that even though GP model and F-theory 
on the elliptically fibered Calabi-Yau on $CP^1\times CP^1$ are in 
the same moduli space, they represent different slices of the 
moduli space. 
In our analysis thus far we did not discover it because the 
deformations (flux of $B$ and possibly $B'$ fields) which take 
us from one slice to the other are neutral under the 
gauge group of the GP model. 

We can proceed further and consider a 
configuration in $F$-theory of the form where:
\be \label{ed4}
h(u,v) = (u-u_1) (v-v_1) P_{3,3}(u,v)\, ,
\ee
where $P_{3,3}$ is a polynomial of degree (3,3) in $(u,v)$. In this case,
near $u=u_1$, $v=v_1$:
\be \label{ed5}
f(u,v)\sim (u-u_1)^2(v-v_1)^2, \qquad 
g(u,v)\sim (u-u_1)^3(v-v_1)^3.
\ee
This corresponds to an intersection of two $D_4$ singularities. 
According to the result of ref.\cite{VAFSIX,BERJOH} the physics of
this situation cannot be described by a local quantum field theory.
In particular we expect to get tensionless strings from three branes
wrapped around the collapsed two cycle.

In the GP model, this configuration corresponds to having two pairs of
branes on top of the orientifold plane at $u=\wt u_1 (=u_1)$, 
two pairs of branes on top of an orientifold plane at $v=\wt v_1(=v_1)$,
and switching off the blow up modes for all the orbifold singularities
lying in the $u=\wt u_1$ and $v=\wt v_1$ plane. But we do not see any
singular behaviour in the GP model at this point in the moduli space.
This mismatch can again be attributed to the presence of the flux of
the $B$ field through the collapsed two cycles. The presence of this
flux breaks the gauge group to $SU(4)\times SU(4)$ due to reasons
outlined before. Also,  as shown in \cite{ASPINK3},
in type IIA theory on an orbifold, presence of this flux prevents a
two brane wrapped around this two cycle to become massless. A simple
argument involving T-duality will tell us that the same mechanism will
prevent a three brane in type IIB theory, wrapped around the two cycle,
to become tensionless. This then is the promised 
explanation of the apparent discrepancy between the F-theory and GP
model results for constant $\lambda$ configuration.

\subsection{Geometry of Intersecting Orientifold Planes and D-branes}

Now that we have established the equivalence of F-theory and the GP
model, we can use the F-theory result to study how non-perturbative
corrections modify the geometry of the intersecting orientifold planes
and D-branes. Of these the fate of intersecting orientifold planes was
already discussed in ref.\cite{PREV} where it was shown that the split
orientifold planes join smoothly near their would be intersection point
to give a pair of complex hyperbola. On the other hand,
since the coupling constant vanishes
on the D-brane, non-perturbative corrections do not modify the geometry
of intersecting D-branes. This is seen from eq.\refb{e22}; the parameter
$C$ that controls the coupling constant does not affect the locations
$u=u_i$ or $v=v_i$ of the D-branes.
However switching on vev of hypermultiplets 
associated with open string states stretched between intersecting D-branes
join a pair of intersecting D-branes into a complex hyperbola as seen from
eq.\refb{e42}. 

It remains to study the fate of intersecting $D$-branes
and orientifold planes. First we consider the intersection of an $A_3$
singularity and an orientifold plane. For this let us consider the
$\Delta$ given in \refb{e22} with $u_1=u_2$,
and analyse the second factor (which gives
the location of the split orientifold plane) near the $A_3$ singularity
$u=u_1$. The locations of the zeroes of $\Delta$ from the second factor
in this approximation is given by
\be \label{ess1}
h^2 - K^2 (u-u_1)^2 = 0\, ,
\ee
for some constant $K$. This can be written as two surfaces:
\be \label{ess2}
h\pm K (u-u_1)=0\, .
\ee
Since this surface is reducible, we see that the
two branches into which the orientifold plane splits intersect at 
$u=u_1$, but do not
join smoothly.

Next we consider intersection of the orientifold plane 
with an $A_1$ singularity. This will correspond
to taking $u_i$, $v_j$ arbitrary in eq.\refb{e22}. Thus now near the 
$A_1$ singularity at
$u=u_1$, the zeroes of the second factor of $\Delta$ are given by
\be \label{ess3}
h^2 - K' (u-u_1) = 0\, ,
\ee
for some other constant $K'$. This surface is not reducible, showing that
the two branches of the split orientifold plane join smoothly at $u=u_1$.

Finally if we destroy the $A_1$ singularity at $u=u_1$ by switching on
more general deformations of the form \refb{e43}, then the factorization
of $\Delta$ into the D-brane part and the orientifold part, as given in
\refb{e20}, is destroyed in general. Thus under such deformation the
D-branes and the two components of the orientifold plane all 
join smoothly near their would be intersection point.

\sectiono{Summary and Conclusion} \label{s4}

In this paper we have established the equivalence between the 
Gimon-Polchinski orientifold, and F-theory on a Calabi-Yau manifold with
Hodge numbers (3,243) by comparaing the gauge symmetry breaking pattern,
local deformations in the moduli space, and background axion-dilaton
fields in the weak coupling limit. It was also found that these two
models represent two different slices of the full moduli space of the
theory. This difference is not visible for most purpose, but becomes
relevant when we analyze subspaces of the F-theory moduli space with
$D_4$ type singularities.

We expect that the techniques used in this paper can be easily generalised
to establish the equivalence between other F-theory - orientifold pairs,
notably in four dimensions. In particular it might be used in establishing
the conjectured duality\cite{GOPMUK}
between the orientifold constructed 
in ref.\cite{BERLEI} and F-theory on an
elliptically fibered Calabi-Yau four fold with base $(CP^1)^3$. 

Our result can also be interpreted as giving non-perturbative information
about the dynamics of a three brane probe on this orientifold. In particular
the background $\lambda$ in F-theory has the interpretation as the
inverse coupling of the U(1) gauge field living on the three brane in the
infrared. The
tree level Lagrangian describing the dynamics of this probe has been 
constructed recently in \cite{COSH}.

\noindent{\bf Acknowledgement}:  I wish to thank R. Gopakumar, D. Jatkar,
S. Kalyanarama, A. Kumar,  S. Mukhi,
J. Sonnenschein and 
S. Yankielowicz for useful discussions. This work was supported in part by a
grant from NM Rothschild and sons Ltd.

\end{document}